\documentclass[aps,superscriptaddress,twocolumn,showpacs,preprintnumbers,amsmath,amssymb,showkeys]{revtex4}
\usepackage[cp1251]{inputenc}
\usepackage[english]{babel}
\usepackage{amssymb,amsmath}
\usepackage{amstext,textcomp} 
\usepackage[dvips]{hyperref}
\usepackage[mathscr]{eucal}
\usepackage{longtable}
\usepackage{graphicx}
\begin{document}
\title{Concept of Classical and Quantum Free Acoustic Field}
\author{Dmitri Yerchuck (a), Felix Borovik (a), Alla  Dovlatova (b), Andrey Alexandrov (b)\\
\textit{(a) Heat-Mass Transfer Institute of National Academy of Sciences of RB, Brovka Str.15, Minsk, 220072, dpy@tut.by\\ 
(b) M.V.Lomonosov Moscow State University, Moscow, 119899}}
\date{\today}%
\begin{abstract} The concept of classical and quantum free acoustic (FA) field is theoretically considered. The equations of the free acoustic field are derived. They coincide in the mathematical form with Maxwell equations for the free electromagnetic (EM) field. It is accentuated, that the equations in the mathematical form of Maxwell equations for the free EM-field are universal equations of the Nature. They describe any free complex-vector  physical field, vector-functions of which are analytical functions. In the case of a free acoustic field, it seems  to be the strong indication taking into account the quantum Fermi liquid model of EM-field leading to the existence of phonons, accompanying the process of a photon formation, that  FA-field and  free EM-field are the display of the single united field, both the components of which can propagate simultaneously [however, with different velocities], for instance, in weakly absorbing  media. The theory proposed can be the theoretical basis for the practical realization of superconducting states at high temperature in the materials with the strong interaction of electrons with acoustic field phonons and with microwave photons in magnetic resonance conditions.\end{abstract}  

\maketitle                         
\section{Introduction}
Corpuscular-wave dualism is referred to fundamental concepts of the modern physics. The given concept is especially clear displayed in the crystals, where there are many fields, displaying both the aspects - wave and corpuscular ones. Quanta of the energy and the quasiimpulse of given fields have obtained the own names. Like that how the term photon describes the corpuscular aspect of an electromagnetic field in  the vacuum, the terms phonon, magnon, polaron, exiton describe some quantised fields in the crystals. Let us remark that  the notion of phonon was introduced in the physics of the crystals by Tamm I.E just by the above indicated analogy with photon. The phonons are considered to be  bounded up with elastic excitations, in particular, the acoustic phonons correspond to usual elastic waves. The quasiimpulse and the energy of phonon are $\vec{q} = \hbar\vec{k}$ and $E = \hbar \omega$ respectively, where $\vec{k}$ and $\omega$ are quasiwave vector and the circular frequency 
of the corresponding wave of the normal lattice vibrations. The vibration state of the crystal in the harmonic approximation can be represented in the form of an ideal phonon gas.

Let us remark that the concept of the phonons is applicable to amorphous solids including glasses for longwave acoustic vibrations, which are weak sensitive to an atomic discrete arrangement and allow the continual description of solids within the frames of the elasticity theory. The concept of the phonons is applicable also to the description of quantum liquids, at that in given case the phonons represent themselves the genuine particles (instead of quasiparticles in solids). Given result was used in the zeroth rest mass boson model of EM-field in \cite{YerchuckDDovlatovaAAlexandrovA} and in the recent work \cite{Yerchuck_Dovlatova_Alexandrov}, in which the EM-field is considered to be quantum liquid and in which the phonon sybsystem is presenting. In fact in the work \cite{Yerchuck_Dovlatova_Alexandrov} the structure of the EM-field vacuum was presented. It consist of massless field bosons for the case of EM-field, photons in which are  excitations - chargeless spin 1/2  topological solitons of Su-Schrieffer-Heeger family.
They were  represented   to be the result of the spin-charge separation effect in the quantum
Fermi liquid and it makes substantially more clear the nature of corpuscular-wave dualism. Really, like to a matter atomic structure, the quantized EM-field represents
itself in the model proposed the discrete "boson-atomic" structure, the individual bosons in which produce the lattice like to a genuine atomic lattice structure in condensed matter. The
main difference between the EM-field "atomic" lattice structure and an atomic lattice structure in condensed matter
consists in that the "atoms" in EM-field structure have
zeroth rest mass. The origin of waves in given structure is determined by the mechanism, quite analogous
to Bloch waves' formation in the solid state of condensed matter. They are harmonic trigonometric functions for
Maxwellian EM-field. At the same time, there
are simultaneously the corpuscules, propagating along
given EM-field "boson-atomic" 1D-lattice structure, that
is, chargeless spin 1/2 topological relativistic solitons -
photons, formed in usual conditions (or spinless charged
solitons in so-called "doped" EM-field structure). Within the frames of the concept proposed it is reasonable to suggest, that the way for the photon propagation [that is EM-field vacuum structure, proposed to be consisting of rest masslees bosons]  was built just after  of the  Sun emergence (and other cosmic light sources) and the velocity of the given process seems to be unknown. The significant consequence of the given model is that the number of massles bosons in vacuum state is finite [being to be the consequence of finite length to light source], that can be used for the development of an electrodynamic theory.

Let us remark, that the model proposed in \cite{YerchuckDDovlatovaAAlexandrovA} and in  \cite{Yerchuck_Dovlatova_Alexandrov} is in fact the development of an idea of multiphoton "molecules". The idea of multiphoton "molecules" arose around 1910, after Debye \cite{Debye} presented  a simple, ondulatory method to obtain the radiation law. Debye idea was developed by using the description of "multiphoton states" in multiphoton "molecules",  in \cite{Wolfke}, \cite{Wolfke_M}, \cite{Boya}: light of definite direction and frequency   presents itself in units "molecules" of 0, 1, 2, ..., n, ...photons, with energy $nh\nu$, that is, with zero binding energy, and they contribute independently to the energy density. The authors of \cite{Boya} accentuate, that the rationale for given assumption is, of course, in the spirit of the atomistic Democritean viewpoint, that all the photons of same frequency are created equal (identical). 
Let us also remark, that the existence of phonons in weakly interacting boson gas media, which can be considered from a quantum liquid viewpoint [that is,  in the media, which mathematically are like to the boson vacuum medium used in the boson model of EM-field in \cite{Yerchuck_Dovlatova_Alexandrov}] is theoretically strictly proved, see, for instance, \cite{Kittel}.

It is substantial for the presented work, that the phonon field in quantum liquids can be considered being to be an independent quantised field. On the other hand, it is well known that any quantised field is a radiation field and it can exist without the sources of its emergence. It means that the free acoustic (FA) field can exist. It can propagate in various media. The given concept was not developed earlier.

 The aim of given work is to  develop  the concept of the FA-field and to give the practical recommendations of its using.

\section{Results and Discussion}

\subsection{Classical  Acoustic Field}

Mathematical characteristics of a classical (that is, not quantised) FA-field   in the general case can be deduced from the analysis of the dynamics of a crystal lattice. It is known that the collective movement of atoms in a crystal lattice can be described by the following  matrix equation 
\begin{equation} 
\label{eqa} 
m\mid\mid\frac{d^2u(\vec{r}_n,t)}{dt^2}\mid\mid = \sum_{n'=1}^{N}\mid\mid\alpha(\vec{r}_n-\vec{r}_n{'})\mid\mid \mid\mid u(\vec{r}_n{'},t)\mid\mid, 
\end{equation}
where $\vec{r}_n$ is the radius-vector of the $n$th lattice atomic site, $m$ is the mass of an  individual atom,  $u(\vec{r}_n,t)$ is the displacement of the $n$th lattice atom from the equilibrium position, $\mid\mid\alpha(\vec{r}_n-\vec{r}_n{'})\mid\mid$ is the crystal dynamic matrix, $N$ is the number of the atoms in a crystal.
The solution of (\ref{eqa}) is 
\begin{equation} 
\label{eqb}  
\mid\mid u_s(\vec{r}_n,t)\mid\mid = \vec{e}_s \mid\mid\exp( i\vec{k}_n\vec{r}_n - i\omega_s t)\mid\mid,
\end{equation}
where $\vec{e}_s $ is the unit vector of the polarization, determining  the direction of the movement of an individual atom, $\vec{k}_n$ is the quasiwave vector,  $\omega_s$ is the circular frequency, which is independent on the atom number, $s \in \overline {1, \nu}$, 
$\nu$ is the number of the atoms in an elementary unit of a crystal.
It is seen from (\ref{eqb}) that the acoustic classical field in crystals is characterised by the complex vector-function. Therefore, in general case the free  acoustic classical field, propgating in a medium, in which the absorption can be neglected can be described by the following vector-function 
\begin{equation} 
\label{eq1} 
\vec{U}_{}(\vec{r},t) = \vec{U}_{1}(\vec{r},t) + i\vec{U}_{2}(\vec{r},t).
\end{equation} 
Let us find the motion equations for the given field. It can be done in a rather simple way, if to take into account that for a free  acoustic  field  the function $\vec{U}_{}(\vec{r},t)$ has to be the analytic function. Then, taking into consideration Cauchy-Riemann conditions, we obtain the following relationships 
\begin{equation}
\label{eq2}
\frac{\partial \vec{U}_{1}(\vec{r},t)}{\partial \vec{r}} = \frac{\partial \vec{U}_{2}(\vec{r},t)}{\partial t}, 
\end{equation}
\begin{equation}
\label{eq3}
\frac{\partial \vec{U}_{1}(\vec{r},t)}{\partial t} = - \frac{\partial \vec{U}_{2}(\vec{r},t)}{\partial \vec{r}}.
\end{equation}
Calculating $\frac{\partial \vec{U}_{1}(\vec{r},t)}{\partial t}$, we have the relation

\begin{equation}
\label{eq4}
\begin{split}
&\frac{\partial \vec{U}_{1}(\vec{r},t)}{\partial \vec{r}} = \\ &(\frac{\partial}{\partial x}\vec{e}_{1} + \frac{\partial}{\partial y}\vec{e}_{2} + \frac{\partial}{\partial z}\vec{e}_{3}) \times\\
&({U}_{1x}(\vec{r},t)\vec{e}_{1} + {U}_{1y}(\vec{r},t)\vec{e}_{2} + {U}_{1z}(\vec{r},t)\vec{e}_{3}) \\
&=-[\nabla \times \vec{U}_{1}(\vec{r},t)]
\end{split}
\end{equation}
Hence
\begin{equation}
\label{eq5}
[\nabla \times \vec{U}_{1}(\vec{r},t)] = -\frac{\partial \vec{U}_{2}(\vec{r},t)}{\partial t}, 
\end{equation}
The analogous calculation for $\frac{\partial \vec{U}_{2}(\vec{r},t)}{\partial t}$ leads to the relation
\begin{equation}
\label{eq6}
[\nabla \times \vec{U}_{2}(\vec{r},t)]= 
\frac{\partial \vec{U}_{1}(\vec{r},t)}{\partial t}  
\end{equation}

Therefore, it is seen, that the equations (\ref{eq5}), (\ref{eq6}) in the mathematical form are coinciding with Maxwell equations. They were obtained from the general conditions, mathematical expressions of  which are appplicable to a very broad class of complex  vector-functions - to  analytical vector-functions. Consequently, they can describe any physical field, wich is described by  analytical complex  vector-functions. Therefore, the equations, which in their mathematical form are coinciding with the equations (\ref{eq5}), (\ref{eq6}) are the univeral equations of the Nature, allowing to describe any physical complex vector field, vector-functions of which are analytical vector-functions. In the case of FA-field considered, it is reasonable to suggest that EM-field in acoustic media is the part of the joint electromagnetic-acoustic field. In other words, speaking in lapidary style, usually well detected EM-field seems to be the only a top of an iceberg. Given suggestion is based along with coincidence of the field equations' form on the concept of the zeroth rest mass boson model of EM-field in \cite{YerchuckDDovlatovaAAlexandrovA} and in \cite{Yerchuck_Dovlatova_Alexandrov}, where the EM-field is considered to be the quantum liquid, in which the phonon sybsystem is presenting.

 Many properties of FA-field and  free EM-field seem to be identical. In particlar, the dual transformations for  FA-field, which  are like to Rainich  dual transformations for  EM-field \cite{Rainich}, will take place. Taking into account the results of the paper \cite{Dovlatova_Yerchuck}, they can be represented in the form
\begin{equation}
\label{eq7}
 \left[\begin{array} {*{20}c}  \vec {U}'_{1}(\vec{r},t) \\ \vec {U}'_{2}(\vec{r},t) \end{array}\right] = \left[\begin{array} {*{20}c} \cos\theta&\sin\theta  \\-\sin\theta&\cos\theta \end{array}\right]\left[\begin{array} {*{20}c}  \vec{U}_{1}(\vec{r},t) \\ \vec{U}_{2}(\vec{r},t) \end{array}\right],
\end{equation}
where $\theta$ is the parameter, the values of which belongs to the segment [0, 2$\pi$].
The relation (\ref{eq7}) can be rewritten
\begin{equation}
\label{eq8}
 \left[\begin{array} {*{20}c}  \vec {U}'_{1}(\vec{r},t) \\ \vec {U}'_{2}(\vec{r},t) \end{array}\right] = e^ {-i\theta}\left[\begin{array} {*{20}c}  \vec{U}_{1}(\vec{r},t) \\ \vec{U}_{2}(\vec{r},t) \end{array}\right],
\end{equation}
It was taken into consideration, that to any matrix, which has the structure, given by the right side in the relation 
\begin{equation}
\label{eq1a}
 f : a + ib \to \left[\begin{array} {*{20}c} a&-b  \\ b&a \end{array}\right]
\end{equation}
 corresponds the complex number, determined by its left side. It is the consequence of the bijectivity of mapping (\ref{eq1a}). In other words, to the real plane determined by orthogonal vectors   $\vec{U}_{1}(\vec{r},t)$, $\vec{U}_{2}(\vec{r},t)$ was set up in the correspondence the complex  plane determined by vectors   $\vec{U}_{1}(\vec{r},t)$, $i\vec{U}_{2}(\vec{r},t)$. It is evident, that    the vectors $\vec {U}'_{1}(\vec{r},t) $ and $\vec {U}'_{2}(\vec{r},t) $ will consist both,  of even  and uneven components under space inversion. So, it is evident from the relation (\ref{eq8}), that if one component of, for instance, $\vec{U}'_{1}(\vec{r},t)$ will be uneven under reflection in the plane, situated transversely to the abscissa-axis, then the second component will be even. 
Invariants of AF if $\theta \neq 0$ are the following
\begin{equation}
\label{eq9}
\begin{split}
&\left[\vec{U}_{1}^2(\vec{r},t) - \vec{U}_{2}^2(\vec{r},t)\right]e^{-2i\theta}\\ 
&+ 2i\left[\vec{U}_{1}(\vec{r},t)\vec{U}_{2}(\vec{r},t)(\vec{r},t)\right] e^{-2i\theta} = inv,
\end{split}
\end{equation}
that is,
\begin{equation} 
\label{eq10}
\begin{split}
\raisetag{40pt}                                                    
&(\vec{U}_{1}^2(\vec{r},t) - \vec{U}_{2}^2(\vec{r},t)) \cos 2\theta +\\ &2(\vec{U}_{1}(\vec{r},t)\vec{U}_{2}(\vec{r},t)) \sin 2\theta = I'_1 = inv \\
&2(\vec{U}_{1}(\vec{r},t)\vec{U}_{2}(\vec{r},t)) \cos 2\theta - \\ &(\vec{U}_{1}^2(\vec{r},t) - \vec{U}_{2}^2(\vec{r},t)) \sin 2\theta = I'_2 = inv.
\end{split}
\end{equation}
 In the case of $\theta = 0$ the FA-field invariants have the form, mathematically coinciding with the well known form for the invariants of the single charge electrodynamics. They are
\begin{equation} 
\label{eq11}
\begin{split}
\raisetag{40pt}                                                    
&(\vec{U}_{1}^2(\vec{r},t) - \vec{U}_{2}^2(\vec{r},t))  = I_1 = inv \\
&(\vec{U}_{1}(\vec{r},t)\vec{U}_{2}(\vec{r},t)) = I_2 = inv.
\end{split}
\end{equation}

\subsection{Classical and Quantized Cavity Acoustic Field}

Suppose an acoustic field in a volume rectangular cavity, which  made up of perfect  walls. Suppose also, that the field is linearly polarized and without any restriction of the  commonness let us choose the one of three possible polarization (the case  $\nu$ = 1 is considered) of the  FA-field component $\vec{U}_{1}(\vec{r},t)$ along $x$-direction. Then the vector-function $\vec{U}_{1}(\vec{r},t)$   can be represented in the form of Fourier sine series

\begin{equation}
\label{eq12}
{U}_{1x}(z,t) \vec{e}_x = \left[\sum_{\alpha=1}^{\infty}A^{{U}_{1}}_{\alpha}q_{\alpha}(t)\sin(k_{\alpha}z)\right]\vec{e}_x , \end{equation}
where
\begin{equation}
\label{eq12a}
k_{\alpha} = \alpha\pi/L; \alpha \in N,
\end{equation}
 $q_{\alpha}(t)$ is the  amplitude of $\alpha$-th normal mode in  the cavity,   $A^{E}_{\alpha}=\sqrt{2 \omega_{\alpha}^2m_{\alpha}/V}$, $\omega_{\alpha} = \alpha\pi c/L$, $L$ is the  cavity length along z-axis, $c$ is sound velocity, $V$ is the cavity volume, $m_{\alpha}$ is the parameter, which is introduced to obtain the analogy with a lattice harmonic oscillator. Let us remember, that the expansion in Fourier series instead of  Fourier integral expansion is  determined by a  discontinuity of $\vec{k}$-space, which is the result of finiteness of cavity volume. Particular sine case of Fourier  series is the consequence of boundary conditions 

\begin{equation}
\label{eq13}
[\vec{n} \times\vec{U}_{1}(\vec{r},t)]|_S = 0, (\vec{n} \vec{U}_{2}(\vec{r},t))|_S = 0,
\end{equation}
which are held true for the perfect cavity considered. Here $\vec{n}$ is the normal to the surface $S$ of the cavity. It can be shown, that $\vec{U}_{1}(\vec{r},t)$ represents itself  a standing wave along z-direction.
For $\vec{U}_{2}(\vec{r},t)$ component, taking into consideration the equations (\ref{eq5}), (\ref{eq6}), we have the expression
\begin{equation}
\label{eq14}
\vec{U}_{2}(\vec{r},t) =  \left[\sum_{\alpha=1}^{\infty}A^{{U}_{2}}_{\alpha}\frac{1}{k_{\alpha}}\frac{dq_{\alpha}(t)}{dt}\cos(k_{\alpha}z) + f_{\alpha}(t)\right]\vec{e}_y,
\end{equation} 
The partial solution, in which the functions $\{f_{\alpha}(t)\}$ are identically zero is

\begin{equation}
\label{eq15}
\vec{U}_{2}^{[1]}(\vec{r},t) =  \left[\sum_{\alpha=1}^{\infty}A^{{U}_{1}}_{\alpha} \frac{1}{k_{\alpha}} \frac{dq_{\alpha}(t)}{dt}\cos(k_{\alpha}z)\right] \vec{e}_y,
\end{equation}
At the same time, there is the second physically substantial solution of  the equations (\ref{eq5}), (\ref{eq6}). Really, from the  expression (\ref{eq14}) for the  field component $\vec{U}_{2}(\vec{r},t)$ differential equations  for
$\{f_{\alpha}(t)\}$, $ \alpha \in N $, can be
 obtained. They are 

\begin{equation}
\label{eq16}
\begin{split}
&\frac{d f_{\alpha}(t)}{dt} + A^{{U}_{1}}_{\alpha}\frac{1}{k_{\alpha}}\frac{\partial^2q_{\alpha}(t)}{\partial t^2}\cos(k_{\alpha}z) \\
&-  A^{{U}_{1}}_{\alpha}k_{\alpha}q_{\alpha}(t)\cos(k_{\alpha}z) = 0.
\end{split}
\end{equation} 
The solutions of (\ref{eq16}) are
\begin{equation}
\label{eq17}
f_{\alpha}(t) =  A^{{U}_{1}}_{\alpha} \cos(k_{\alpha}z)\left[k_{\alpha} \int\limits _{0}^{t} q_{\alpha}(\tau)d\tau -\frac{dq_{\alpha}(t)}{dt}\frac{1}{k_{\alpha}}\right]
\end{equation}
Then $\vec{U}_{2}(\vec{r},t)$ component can be represented in the form
\begin{equation}
\label{eq18}
\vec{U}^{[2]}_{2}(\vec{r},t) = -\left\{\sum_{\alpha=1}^{\infty} A^{{U}_{2}}_{\alpha} q_{\alpha}'(t)\cos(k_{\alpha}z) \right\}\vec{e}_y,
\end{equation}
where $A^{{U}_{2}}_{\alpha}=\sqrt{2 \omega_{\alpha}^2m_{\alpha}/V}$.
Similar consideration gives the second  solution for $\vec{{U}_{1}}(\vec{r},t)$ AF component 
\begin{equation}
\label{eq19}
\vec{{U}_{1}}^{[2]}(\vec{r},t) = \left\{\sum_{\alpha=1}^{\infty} A^{{U}_{1}}_{\alpha}q_{\alpha}''(t)\sin(k_{\alpha}z)\right\}\vec{e}_x,
\end{equation}

 The functions $q_{\alpha}'(t)$ and $q_{\alpha}''(t)$ in (\ref{eq18}) and (\ref{eq19})  are
\begin{equation}
\label{eq20}
\begin{split}
&q_{\alpha}'(t) = {\omega_{\alpha}}\int\limits _{0}^{t} q_{\alpha}(\tau)d\tau\\
&q_{\alpha}''(t) = {\omega_{\alpha}}\int\limits _{0}^{t} q_{\alpha}'(\tau')d\tau'
\end{split}
\end{equation}
The class of field functions $\{q_{\alpha}(t)\}$ has to satisfy to the equations (\ref{eq5}), (\ref{eq6}). Given condition allow to find the given class. It will satisfy to  differential equations
\begin{equation}
\label{eq5a}
\frac{d^2q_{\alpha}(t)}{dt^2} + {k_{\alpha}^2} q_{\alpha}(t)= 0, \alpha \in N.
\end{equation}
Consequently,  we have
\begin{equation}
\label{eq6a}
q_{\alpha}(t) = C_{1\alpha} e^{i\omega_{\alpha}t} + C_{2\alpha} e^{-i\omega_{\alpha}t}, \alpha \in N,
\end{equation}
where  $C_{1\alpha}, C_{2\alpha}, \alpha \in N$, are arbitrary constants.
Thus, the solutions are complex-valued functions, although the variables in free acoustic field differential equations are real-valued vector-functions. The given  result is  well known in the theory of differential equations.  
It means, that generally the field functions for  the free acoustic field in the cavity produce complex space.  On the other hand, the equation (\ref{eq5a}) has also the only real-valued general solution, which can be represented in the form
\begin{equation}
\label{eq6ab}
q_{\alpha}(t) = B_{\alpha} \cos(\omega_{\alpha}t + \phi_{\alpha}),
\end{equation}
where  $B_{\alpha}, \phi_{\alpha}, \alpha \in N$ are arbitrary constants. 

So, for the classical cavity FA-field case we can in the principle restrict themselves to the only real-valued general solution of FA-field equations.

Owing to the fact, that the solutions of FA-field equations have the simple form of harmonic trigonometrical functions, it is easily  to establish that the 
second solution for acoustic  field functions differs from the first solution the only by sign, that is substantial,  and by inessential integration constants.  Integration constants  can be taken into account by means of redefinition of factor $m_{\alpha}$ in field amplitudes.  

Let us briefly consider the quantization of free acoustic field.
We  begin the consideration  like to the canonical quantization, from classical Hamiltonian, which for the first partial classical solution of field equations is
\begin{equation}
\label{eq21}
\begin{split}
&\mathcal{H}^{}(t) = \frac{1}{2}\int\limits_{(V)}\left[\vec{U}_{1}^2(z,t)+\vec{U}_{2}^2(z,t)\right]dxdydz\\
&= \frac{1}{2}\sum_{\alpha=1}^{\infty}\left[m_{\alpha}\omega_{\alpha}^2q_{\alpha}^2(t) + \frac{p_{\alpha}^2(t)}{m_{\alpha}} \right],
\end{split}
\end{equation}
where
\begin{equation}
\label{eq22}
p_{\alpha} = m_{\alpha} \frac{dq_{\alpha}(t)}{dt}.
\end{equation}
We set up then in the correspondence to canonical variables ${q}_{\alpha}(t), {p}_{\alpha}(t)$, determined by the first partial solution of field equations,  the operators by a usual way
\begin{equation}
\label{eq23}
\begin{split}
&\left[\hat {p}_{\alpha}(t) , \hat {q}_{\beta}(t)\right] = i\hbar\delta_{{\alpha}\beta}\\
&\left[\hat {q}_{\alpha}(t) , \hat {q}_{\beta}(t)\right] = \left[\hat {p}_{\alpha}(t) , \hat {p}_{\beta}(t)\right] = 0,
\end{split}
\end{equation}
where
$\alpha, \beta \in N$.
Let us define the operator functions  of  the time
$\hat{a}_{\alpha}(t)$ and $ \hat{a}^{+}_{\alpha}(t)$
\begin{equation}
\label{eq24}
\begin{split}
&\hat{a}_{\alpha}(t) = \frac{1}{ \sqrt{ 2 \hbar  m_{\alpha} \omega_{\alpha}}} \left[ m_{\alpha} \omega_{\alpha}\hat {q}_{\alpha}(t) + i \hat {p}_{\alpha}(t)\right]\\
&\hat{a}^{+}_{\alpha}(t) = \frac{1}{ \sqrt{ 2 \hbar  m_{\alpha} \omega_{\alpha}}} \left[ m_{\alpha} \omega_{\alpha}\hat {q}_{\alpha}(t) - i \hat {p}_{\alpha}(t)\right].
\end{split}
\end{equation}
Then the operator functions of canonical variables can be represented in the form
\begin{equation}
\label{eq25}
\begin{split}
&\hat {q}_{\alpha}(t) = \sqrt{\frac{\hbar}{2 m_{\alpha} \omega_{\alpha}}} \left[\hat{a}^{+}_{\alpha}(t) + \hat{a}_{\alpha}(t)\right]\\
&\hat {p}_{\alpha}(t) = i \sqrt{\frac{\hbar m_{\alpha} \omega_{\alpha}}{2}} \left[\hat{a}^{+}_{\alpha}(t) - \hat{a}_{\alpha}(t)\right]. 
\end{split}
\end{equation}
AF  operator functions are obtained taking into account (\ref{eq25}) right away and they are
\begin{equation}
\label{eq26}
\hat{\vec{U}}_{1}(\vec{r},t) = \{\sum_{\alpha=1}^{\infty} \sqrt{\frac{\hbar \omega_{\alpha}}{V}} \left[\hat{a}^{+}_{\alpha}(t) + \hat{a}_{\alpha}(t)\right] \sin(k_{\alpha} z)\} \vec{e}_x,
\end{equation}
and

\begin{equation}
\label{eq27}
\hat{\vec{U}}_{2}(\vec{r},t) = i \{\sum_{\alpha=1}^{\infty} \sqrt{\frac{\hbar \omega_{\alpha}}{V}} \left[\hat{a}^{+}_{\alpha}(t) - \hat{a}_{\alpha}(t)\right] \cos(k_{\alpha} z)\} \vec{e}_y,
\end{equation}
Taking into account the relationships (\ref{eq26}), (\ref{eq27}) and AF equations (\ref{eq5}), (\ref{eq6}), we find  an explicit form for the dependencies of operator functions   $\hat{a}_{\alpha}(t)$ and $ \hat{a}^{+}_{\alpha}(t)$ on the time. They are the following

\begin{equation}
\label{eq28}
\begin{split}
&\hat{a}^{+}_{\alpha}(t) = \hat{a}^{+}_{\alpha}(t = 0) e^{i\omega_{\alpha}t},\\
&\hat{a}_{\alpha}(t) = \hat{a}_{\alpha}(t = 0) e^{-i\omega_{\alpha}t},
\end{split}
\end{equation}
where $\hat{a}^{+}_{\alpha}(t = 0), \hat{a}_{\alpha}(t = 0)$ are constant operators.
Physical sense of operator time dependent functions $\hat{a}^{+}_{\alpha}(t)$ and $\hat{a}_{\alpha}(t)$ is understandable. They are creation  and annihilation operators of the $\alpha$-mode phonon in the cavity.
They are continuously differentiable operator functions of  a time. It means, that the time of  phonon creation (or annihilation) can be determined strictly, at the same time operator  functions $\hat{a}^{+}_{\alpha}(t)$ and $\hat{a}_{\alpha}(t)$  do not curry any information on the place, that is on space coordinates of a given event. 

Thus, we have considered briefly the local time quantization of  a free acoustic field.  The local space quantization and the local space-time quantization of  a free acoustic field can be made in a similar way, which was developed for the local space quantization and the local space-time quantization of a cavity EM-field in \cite{Dovlatova_Yerchuck}. The possibility to quantize free AF means that free AF is really the radiation field and it can exist without the sources of its initial emergence.

It seems to be also essential, that complex exponential dependencies in (\ref{eq28}) cannot be replaced by the real-valued harmonic trigonometrical functions. The proof is identical to the proof of an analogous statement for Maxwellian EM-field \cite{Dovlatova_Yerchuck}. Consequently, the quantized free acoustic field differs qualitatively  from the classical acoustic  field.

Practical significance of the results above presented is the following. By the study of the properties of the solids or quantum liquids [for the subsequent using in engeneering] situated in the cavity [for instance, by radio or optical spectroscopy methods] we have to take into account that two discrete $\vec{k}$ spaces will be formed with different discretness extent, determined by interatomic spacing in solids  on the one hand and  by cavity modes on the second hand. At that, they can be incommensurable. In given case, the cavity FA-field phonons can lead by the interaction with electronic subsystem of solids practically the only to exchange by the energy and the impulse between electrons [that is, to the formation of Cooper pairs]. 
Really, the processes of electron scattering with the generation or the absorption of FA-field phonons are impossible, since the probability of the processes of electron scattering with the generation or the absorption of phonons, for instance, for nondegenerated semiconductors is determined by  the known expression
\begin{equation}
\label{eq29}
W^{\pm\vec{q}_j}_{\vec{k}_l\rightarrow\vec{k}_m} = \frac{2\pi}{\hbar}{\mid M^{\pm\vec{q}_j}_{\vec{k}_l\rightarrow\vec{k}_m}\mid}^2 \delta(\mathcal{E}_{\vec{k}_l}-\mathcal{E}_{\vec{k}_m} \mp \hbar\omega_{\vec{q}_j})[N_{\vec{q}_j} + \frac{1}{2} \pm \frac{1}{2}],
\end{equation}
upper signs in which correspond to the phonon emission, lower signs  correspond to the phonon absorption.
Here $\vec{k}_l$, $\vec{k}_m$ are the  quasiwave vectors of the lattice, $l,m = 1, 2, ..., N'$, determined by discrete structure of $\vec{k}$ lattice space with the discretness
\begin{equation}
\label{eq30}
 k_n = \frac{2\pi n}{N'}, n = 1, 2, ..., N', 
\end{equation}
 $N'$ is the number of the atoms in the lattice, $\mathcal{E}_{\vec{k}_l}$ and $\mathcal{E}_{\vec{k}_m}$ are the values of the energy of the electron in the states with  quasiwave vectors $\vec{k}_l$ and $\vec{k}_m$   correspondingly, $\vec{q}_j$ is quasiwave vector of absorbed or emitted phonon of the lattice with the value $q_j$, satisfying to the  relation (\ref{eq30}), by the replacement $k\rightarrow q$, $\omega_{\vec{q}_j}$ is circular phonon frequency, $N_{\vec{q}_j}$ is a number of phonons with  quasiwave vector $\vec{q}_j$,
$M^{\pm\vec{q}_j}_{\vec{k}_l\rightarrow\vec{k}_m}$ is matrix element of the transition $\vec{k}_l\rightarrow\vec{k}_m$, for which the conservation law of the quasiimpulse, satysfying to the relation
\begin{equation}
\label{eq31}
\begin{split}
\hbar\vec{k_l}- \hbar\vec{k_m} \mp \hbar\vec{q}_j = \vec{b}
\end{split}
\end{equation}
takes place. Here $\vec{b}$ is an arbitrary vector of a reciprocal lattice. Let us remember, that the thansitions with $\vec{b} = 0$ are normal transitions, if $\vec{b} \neq 0$, the thansitions are accompanying with Peierls transfer processes. It can be concluded, that FA-field phonons will be not satisfy in the general case to the expression (\ref{eq29}) and consequently they will not contribute to scattering processes. It follows from that that they will have the quasiimpulse distribution obeing to the   relation (\ref{eq12a})
in  which $\vec{k}$-space discretness is determined by the cavity length in the microwave field propagation direction. It is understandable that in the case of incommensurate cavity and lattice $\vec{k}$ space discretnesses all the generated hypersound phonons will have the energy and quasiimpulse characteristics, which will be not satisfying to the relation (\ref{eq29}). Let us remark, that the mechanism of  the creation of the system of coherent resonance hypersound phonons is described in the case  of the strong interaction of electrons with cavity FA-field phonons and cavity photons (the term "strong" means that practically all conducting electrons in a sample are involved in the given process directly or undirectly and they will not participate in the scattering processes with equilibrium own lattice phonons, see further)
in the work \cite{Alla}.

Simultaneously, the processes of electron scattering with the emission or the absorption of own lattice equilibrium phonons will be suppressed in the case considered. Really, a photon flux  incoming in the cavity in the microwave region is very intensive. So, the photon flux, corresponding to the microwave power in 100 mW, that is to the standard power of microwave generators in ESR-spectrometers without an attenuation, is equal to $\approx 10^{22}$ photons/s  of the frequency equaled to 10 GHz. It leads by a strong electron-photon interaction, realised by a spin subsystem of a sample electron system, to a very great number of cavity FA-field hypersound phonons.
At the same time, it is well known, that the  interaction which produces the energy difference
between the normal and superconducting phases in theory of superconductivity
 arises from the virtual exchange of phonons and
the screened Coulomb repulsion between electrons.
Other interactions are thought to be essentially
the same in both states, their effects thus cancelling in
the energy difference.
The Hamiltonian for the the electron-phonon interaction,
which comes from virtual exchange of phonons between
the electrons is the following \cite{BCS}
\begin{equation}
\label{eq32}
\begin{split}
&\mathcal{H}_{e-ph} = \\
&\hbar\sum_{\vec{k}\vec{k'}\sigma\sigma{'} \vec{q}}\frac{\omega_{\vec{q}}|M_{\vec{q}}|^2\hat{c}^{+}(\vec{k'}-\vec{q},\sigma{'})\hat{c}(\vec{k'},\sigma{'})\hat{c}^{+}(\vec{k}+\vec{q},\sigma)\hat{c}(\vec{k},\sigma)}{(\varepsilon_{\vec{k}} - \varepsilon_{\vec{k}+\vec{q}})^2 - (\hbar\omega_{\vec{q}})^2 }
\end{split}
\end{equation}
where $\varepsilon_{\vec{k}}$ is the Bloch energy measured relative to the
Fermi energy, $M_{\vec{q}}$ is matrix element for phonon-electron
interaction, $\hat{c}^{+}(\vec{k},\sigma)$, $\hat{c}(\vec{k},\sigma)$ are fermion  creation and
annihilation operators, based on the renormalized
Bloch states specified by quasiwave vector $\vec{k}$ and spin projection $\sigma$,
which satisfy the usual Fermi commutation relations. 
The most significant transitions are those for which  $|\varepsilon_{\vec{k}} - \varepsilon_{\vec{k}+\vec{q}}| \ll (\hbar\omega_{\vec{q}}$. It is evident, that by the generation of the coherent system of resonance hypersound phonons the given condition will be easily fulfilled. The probality of given process is proportional to $\sum_{\vec{q}}$ in the expression (\ref{eq32}), that is, to the number of FA-field hypersound phonons wich can be produced in a very large amount. It gives clear proof for the participation of the conducting electron system in the materials with the strong electron-phonon and electron-photon interactions at ESR conditions the only in the virtual exchange processes, excluding the scattering processes.

 At the same time it is well known, that just the electron scattering  processes are the main processes determining the electrical resistance. Thus, it is the direct way for the formation of the superconducting states  in ESR conditions or another magnetic resonance conditions by room temperature and even by more high temperatures in conductors or semiconductors in which  the strong interaction of sample electrons with cavity FA-field phonons  is realized.

 The given conclusion has the  experimental confirmation.   Peculiarities of spin-wave resonance observed in carbon nanotubes, produced by high-energy
ion beam modification of diamond single crystals in $\langle{100}\rangle$ direction \cite{Dmitri} 
allowed to insist on the formation in given nanotubes of
s+-superconductivity at room temperature, coexisting with
uncompensated antiferromagnetic ordering just in ESR conditions. The role of hypersound resonance phonons produced in ESR-spectrometer cavity, that is the phonons which can be referred to cavity FA-field quanta, in the formation of the superconducting state in a microwave frequency range is proposed in the cited work to be crucial. 

\section{Conclusions}

The concept of classical and quantum free acoustic field is theoretically considered for the first time. The equations of the free acoustic field are derived. They coincide in the mathematical form with Maxwell equations for a free electromagnetic field. It is concluded, that the equations, which in their mathematical form are coinciding with the equations (\ref{eq5}), (\ref{eq6}) are the univeral equations of the Nature.  They allow to describe any physical complex vector field, vector-functions of which are analytical vector-functions. 

It is proposed, taking into account the quantum Fermi liquid model of an electromagnetic field leading to the existence of phonons, accompanying the process of a photon formation, that the same form of field equations is the strong indication, that both the free acoustic field and the free electromagnetic field aew the display of the single united field, both the components of which can propagate simultaneously [however, with different velocities], for instance, in weakly absorbing  media. 

The theory of the free acoustic field can be the theoretical basis for the practical realization of superconducting states at high temperature in the materials with the strong interaction of electrons with acoustic field phonons and microwave photons in magnetic resonance conditions. The given conclusion has the  experimental confirmation for the case of electron spin resonance, which is briefly described in the paper presented.


\begin{thebibliography}{16}

\bibitem{YerchuckDDovlatovaAAlexandrovA} Dmitri Yerchuck, Alla Dovlatova, Andrey Alexandrov,  ICNT Conference, Paris, France, 2012, Book of Abstracts, PO3.5
\bibitem{Yerchuck_Dovlatova_Alexandrov} Yerchuck Dmitri, Dovlatova Alla, Alexandrov Andrey, to be published
\bibitem{Debye} Debye P, Ann.d.Phys.,  \textbf{33} N 4 (1910) 1427-1434
\bibitem{Wolfke}	Wolfke M, Zur Quantentheorie, Verh.DPG15 (1913) 1123-1129, 1215-1218
\bibitem{Wolfke_M} Wolfke M, Phys.Zs.,\textbf{22} (1921) 375-379
\bibitem{Boya}		Boya Luis J, Duck Ian M, and Sudarshan E C G, arXiv:quant-ph/0010010v1 2 Oct (2000)
\bibitem{Kittel} Kittel C, Quantum Theory of Solids, John Wiley and Sons Inc., New York - London, 1963, M., Nauka, 1967, 492 pp   
\bibitem{Rainich} Rainich G Y, Trans.Am.Math.Soc.,\textbf{27} (1925) 106–136
\bibitem{Dovlatova_Yerchuck} Alla Dovlatova and Dmitri Yerchuck, Journal of Physics: Conference Series, \textbf{343} (2011) 012133, DOI:10.1088/1742-6596/343/1/012133 
\bibitem{Dmitri} Dmitri Yerchuck, Yauhen Yerchak, Vyacheslav Stelmakh, Alla Dovlatova, Andrey Alexandrov, J.Supercond.Nov.Magn.,(2013) DOI 10.1007/s10948-013-2300-7
\bibitem{Alla} Alla Dovlatova, Dmitri Yerchuck, ISRN Optics, Volume 2012, Article ID 390749, 10 pages, DOI:10.5402/2012/390749
\bibitem{BCS} Bardeen J, Cooper L N, Schrieffer J.R, Phys.Rev., \textbf{108}, N 5 (1957)
1175–1204 
\end{thebibliography}
\end{document}